\documentclass[a4paper, amsfonts, amssymb, amsmath, reprint, showpacs, showkeys, nofootinbib, twoside, floatfix]{revtex4-1}
\usepackage[english]{babel}
\usepackage[utf8]{inputenc}
\usepackage{upgreek}
\usepackage[colorinlistoftodos, color=green!40, prependcaption]{todonotes}
\usepackage{amsthm}
\usepackage{mathtools}
\usepackage{physics}
\usepackage{xcolor}
\usepackage{graphicx}
\usepackage[left=16mm,right=16mm,top=35mm,bottom=20mm,columnsep=14pt]{geometry} 
\usepackage{adjustbox}
\usepackage{placeins}
\usepackage[T1]{fontenc}
\usepackage{lipsum}
\usepackage{csquotes}
\usepackage{url}
\usepackage{natbib,hyperref}
\usepackage{cleveref}
\hypersetup{colorlinks=true, urlcolor=blue, linkcolor=blue, citecolor=blue}
\usepackage{multirow, booktabs}
\usepackage{epstopdf}

\begin{document}

%\preprint{APS/123-QED}

\title{Ion-pair dissociation dynamics in electron collision with carbon dioxide probed by velocity slice imaging}% Force line breaks with \\

\author{Narayan Kundu}
\author{Sumit Naskar} 
\author{Irina Jana} 
\author{Anirban Paul}
% \altaffiliation[Also at ]{Indian Institute of Science Education and Research (IISER), Kolkata}%Lines break automatically or can be forced with \\
\author{Dhananjay Nandi}%
 \email{dhananjay@iiserkol.ac.in}
\affiliation{Indian Institute of Science Education and Research Kolkata, Mohanpur 741246, India}
%\affiliation{$^\dagger$Present address: Radiation Laboratory, University of Notre Dame, Notre Dame, Indiana 46556, United States}%


%\affiliation{
%\affiliation{
%\affiliation{%



\begin{abstract}

Ion-pair dissociation (IPD) to gas phase carbon dioxide molecule has been studied using time of flight (TOF) based mass spectroscopy in combination with the highly differential velocity slice imaging (VSI) technique. The appearance energy of the fragmented $O^{-}$ ion provides the experimental threshold energy value for the ion-pair production. The kinetic energy (KE) distributions and angular distributions (AD) of the fragment anion dispense the detailed insight to the IPD dynamics. The KE distribution clearly reveals that the IPD dynamics may be due to the direct access to the ion-pair states. However, indirect mechanism can't be ruled out at higher incident electron energies. The angular distribution data unambiguously identified the involvement of ion-pair state associated with $\Sigma$ symmetry and minor contribution from $\Pi$ symmetric states. Computational calculations using density functional theory (DFT) strongly support the experimental observations.

%\pacs{34.50.Gb, 34.80.Ht}

%\end{description}
\end{abstract}

\maketitle

%\tableofcontents

\section{Introduction}

In recent years dynamics study with anion fragments has marked a new realm in the field of electron induced chemistry \cite{nikjoo1997computational,boamah2014low}. Dissociative electron attachment (DEA) and ion-pair dissociation (IPD) are the two well established inelastic scattering phenomena to probe such anion dynamics in low (0 - 15 eV) and  intermediate (15 - 50 eV) electron molecule collisions, respectively. In DEA, resonances occur at some particular energies (less than ionization energy of that particular molecule) whereon the colliding particles generate quasi-bound temporary negative ion (TNI) complexes \cite{whelan2006electron}; resulting in rapid variations in the energy dependence of the inelastic  scattering cross sections. Contrastingly, IPD is a non-resonant post-DEA phenomenon in ionization continuum. In general, ion-pair dissociation originates via. some well defined Rydberg states, even if direct dissociation to an ion-pair states provides a finite probabilistic value for some cases \cite{suits2006ion, chakraborty2016dipolar}. However, the energetically higher (greater than ionization potential energy of that particular molecule) neutral states in ionization continuum are responsible for such ion-pair generation. These well defined high energetic Rydberg states are commonly known as super-excited states \cite{Kouchi2013superexcited,kouchi1997dissociation,hao2017superexcited}. During the study of interaction process between ionizing radiation and  matter, Platzman first gives the importance of the super-excited states \cite{platzman1962superexcited}. The dissociation dynamics obtained from these super-excited states varies greatly with that of obtained from the ordinary Rydberg states excited below ionization thresholds \cite{hatano2001interaction}. Suitable amount of energy transferred to the target molecule using electron or photon collisions reveals this type of states to play with. IPD often occurs when the incident electron partially transfers its kinetic energy to the molecule and leaves the molecule to one of its neutral super-excited states or neutral ion-pair states. Suppose, the molecule AB inelastically scatters due to  electron molecule interaction and reaches the states beyond its ionization. Then the future of this energy pick up by the molecule will be the reason for achieving one of the following processes. 
\begin{align}
    \rm{AB} + \ energy \ & \longrightarrow  \rm{AB}^{+}  + e^{-} \ : \ direct \ \ ionization\\
                    & \longrightarrow \rm{AB}^{**}  \ : \ super \ excitation
\end{align}
  Finally, the super excitation gives us phenomena like:-
 \begin{align}
     \rm{AB}^{**} & \longrightarrow \rm{AB}^{+}  + e^{-}  : \ \rm{autoionization}\\
             & \longrightarrow \rm{A}^{+}  + \rm{B}^{-}   : \ \rm{ion-pair \ dissociation}\\
             & \longrightarrow \rm{A} + \rm{B} \ : \ \rm{neutral \ dissociation}\\
             & \longrightarrow \  \rm{fluorescence\ and \ other\  processes}
 \end{align}
The ion-pair threshold not only depends on geometry of that ion-pair states but also on the geometry of the target and products as well; resulting in the threshold energy value to be channel specific.

In case of collision between photon and isolated molecules, coherent vacuum ultraviolet laser sources play a crucial role for obtaining the IPD dynamics in a similar fashion like photoelectron spectroscopy. Threshold ion pair production spectroscopy (TIPPS) \cite{shiell2000threshold, martin1998determination} and ion pair imaging spectroscopy (IPIS) \cite{yang2005combined,suto1997ion} are the two traditionalistic approaches to probe such ion-pair states. TIPPS is focused at around the dissociation threshold where as IPIS is based on ion imaging of the energetic fragments. Although IPD process is most studied using photon impact, the electron collision with isolated molecules can provide us similar types of results as obtained from the techniques mentioned below. Thus, the super-excitation as well as direct excitation to the ion pair states can be investigated by the electron collision. During the dynamics studies of DEA or IPD products, this VSI probing technique has satisfactorily provided us adequate information over the last two decades \cite{nandi2005velocity, ram2012dissociative, chakraborty2016dipolar}. 

In the present work, we apply the VSI technique in a TOF based mass spectrometry to obtain the IPD pathway for a triatomic molecule: carbon dioxide (CO$_2$). Here, the energetic and dynamic informations are reported for the wide range of incident electron-energy spanning roughly from 20 to 40 eV. Till today, there is no experimental report about the threshold energy value of incident electron to kick off ion-pair origination in CO$_2$. Kinetic energy (KE) and angular distributions (AD) of the fragmented anion are also unknown to the scientific community for the so-called molecule. In this article, our work have been focused on the study of both KE and AD for produced O$^-$ channel along with ion-pair threshold behaviour using the high resolution VSI spectroscopy. 

\section{Instrumentation}

The experimental set up used for this study is based on crossed-collision between electron and molecular beam and has already been discussed in details elsewhere \cite{nag2015complete}. The measurements are done by the highly differential time sliced velocity map imaging (VMI) spectrometer specially built to study low energy electron-molecule interactions. The experimental set-up is made of with a magnetically collimated and pulsed electron beam, a Faraday cup, effusive molecular beam, a three field time-of-flight velocity map imaging electrode system, and a two-dimensional position sensitive (PSD) detector with linked data acquisition system. The experiment is carried out in a vacuum chamber, which is pumped down up-to $10^{-10}$ mbar using a turbo molecular vacuum pump and scroll pump arrangement. In this experiment, the pulsed electron beam interacts with an effusive molecular beam formed by a capillary tube. After finite time delay, the fragmented anions (O$^-$)are extracted into a time-of-flight mass spectrometer using a suitable extraction pulse.

Thermionic emission process is applied successfully to generate our electron beams whereon a programmable power supply (Instek PSM-6003) controls the beam energy. The energy resolution of this electron beam is about 0.6 to 0.7 eV. The electron beams used here are pulsed in nature. The typical pulse width is about 200 nanosecond with 10 kHz repetition rate. A Faraday cup measures the time averaged current using the collected electrons. Our TOF based three field configuration spectrometer capable of mapping the fragmented anions on the two dimensional position sensitive detector. All the anions created with a particular velocity vector are mapped onto a single point on the 2D PSD, consisting of three microchannel plates (MCP) with Z-stack configuration. A three-layer delay line hexanode is used to encode position information \cite{jagutzki2002multiple} for the detected anions. The TOF of the detected ions are calculated from the back-MCP signals. The position informations (x and y) for the detected ions are redeemed using the hexanode placed behind the MCPs. The flight time (t) and the redeemed position (x, y) of each detected ion is saved in a list-mode format (LMF) generated through CoboldPC software \cite{ullmann1999list}. The experiments are performed under ultra-high vacuum conditions having pressure $10^{-7}$ mbar. $ 99.9 \%$ pure commercially available CO$_2$ gas is used to perform this experiment. Only the MCP output is used to obtain the ion yield curve. In this case, a fast amplifier amplifies the MCP signal and then sent to a constant fraction discriminator (CFD). CFD output signal act as a STOP signal for nuclear instrumentation module (NIM) that converts the time signal to a corresponding amplitude, commonly known as time to amplitude converter (TAC). A master pulse generator used to generate the START pulse synchronized with the electron gun pulse. The TOF of the detected O$^-$ ion is nothing but time difference between this START and STOP signals. The TAC output signal is connected to a multichannel analyzer (MCA, Ortec model ASPEC-927). Using USB 2.0 port, MCA output signal is transferred to a computer for further analysis. Lastly, the mass spectra and the ion yields for the fragments are measured using LABVIEW based data acquisation system.

When the electrons with intermediate energy (15 to 50 eV) interact with the molecule, each ion-pair dissociation event yields two partner fragments flying with equal momentum in opposite directions in the centre of mass frame. When we repeat the same event for many times, spherical distributions of fragments will be build up in velocity space. These spherical distributions of fragments are the so called `Newton (velocity) Sphere's. The size of this Newton Sphere tells us about the balance of internal and translational energy in the reaction \cite{whitaker2003imaging}. One can extract the kinetic energy and  angular distribution statistics of the fragments from the projection of the Newton Sphere onto a position sensitive detector. Here the radius of a particular Newton Sphere is proportional to the kinetic energy of that particular fragment. Thus, the fragmented ions with higher momentum must fall onto the PSD with bigger diameter.

In this present experiment, 4 micro-seconds extraction pulse is used. This extraction pulse is functionally active in such a time scale that fragments get enough time to construct the Newton sphere. Usually we have taken 100 nanosecond delay with respect to electron gun pulse for starting the extraction pulse. This approach not only allow to achieve better resolution for time sliced images but also restrict the incident electrons to fall onto the position sensitive detector. One can obtain the slice images from the detected Newton Sphere using suitable time window. The CoboldPC from RoentDek is used to slice the raw data. The central is more informative due to its biggest diameter as well as formed in electron beam plane territory. Thus, the central time slice images are taken into consideration to analyse angular distribution (AD) for the fragmented ions (usually anions). So, the fragmented ions for central slice are roughly parallel to the detector. Generally, the fragmented O$^-$ ions detected by 2D PSD lie within the full with half maximum (FWHM) about 500 nanoseconds. To determine to kinetic energy of the fragmented ions, Kinetic energy calibration is performed using the well known DEA resonant peaks for O$^-$ ions produced due to electron collision with gas phase oxygen moledcule.

\section{Results and Discussions}

\begin{figure}[hbt!]
       \centering
       \includegraphics[scale=0.15]{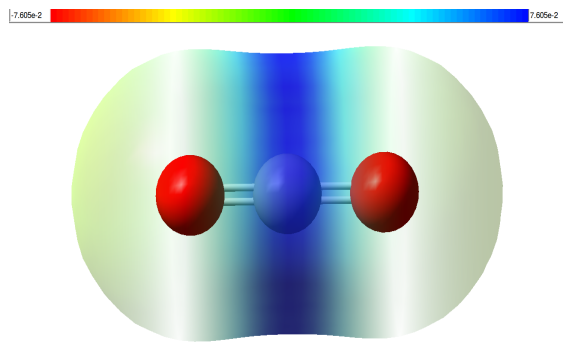}
       \caption{Electrostatic potential map at 0.004 electron/Bohr$^3$ iso-density surface of CO$_2$ molecule. The red in the color spectrum denotes lowest electrostatic potential value while the blue denotes the highest. The red spheres represent the O-atoms and the blue sphere represents the C-atom.}
       \label{fig:emp}
\end{figure}

If we consider the collision between gas phase CO$_2$ and electrons with intermediate energy, then produced intermediate super-excited state may decay through the channels as discussed below.
\begin{equation}
     \rm{CO}_2 + e^- \longrightarrow CO_2^{**} + e^-\longrightarrow
\begin{cases}
     \rm{CO}^{+} + O^- + e^-\\
     \rm{CO}^{-} + O^+ + e^-
\end{cases}
 \end{equation}
Though CO$_2$ is a non-polar molecule because of the symmetry of its bonding, the double covalent bond attached with each oxygen atom is polar in nature. When the bond dissociation of CO$_2$ comes into picture, the charge cloud generated through this molecule’s electrons and nuclei can serve as fate to ingress the molecule’s reactivity towards ion-pair formation. Thus, three dimensional visualization of molecular charge distribution i.e. electrostatic potential (ESP) map will play a key role to rationalize the most probabilistic ion-pair generation channel. Considering B3LYP functional with 6-31G(d,p) basis set, an $ab\ \ initio$ electronic structure calculation is performed using density-functional theory (DFT) to optimize the geometry of this molecule. The optimized systemic framework is used to plot the ESP map at the  ground state of CO$_2$ as shown in Fig. \ref{fig:emp}.
   
\begin{figure}[hbt]
      \centering
      \includegraphics[scale=0.34]{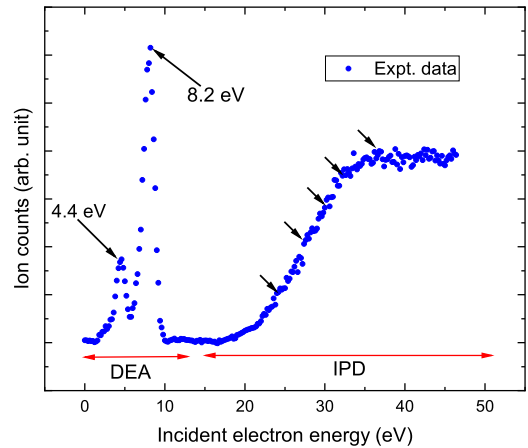}
      \caption{The data points in blue color represent the number of produced anion (O$^-$) as a function of electron beam energy. Both DEA and IPD region are shown by red arrows. Black arrows represents the incident beam energies accounted for studying IPD dynamics.}
      \label{fig:ionyield}
\end{figure}

From the Fig. \ref{fig:emp}, it is obvious to state that the C$=$O bond is strongly polarised towards oxygen due to higher electro-negativity of oxygen atom as compared to carbon atom. So, we can easily vindicate that the IPD of CO$_2$ is more likely via. first channel i.e. the formation of O$^{-}$ and CO$^{+}$. Again, the variation in the number of ion-pair formations is plotted as the function of incident electron beam energy, conventionalised ion yield curve as shown in Fig. \ref{fig:ionyield}. The most conspicuous influences on the ion yield characteristic are electron affinity for negative ions and ionization potential energy for positive ion. This ion yield curve is in good agreement with the DEA-resonances at 4.4 and 8.2 eV incident electron energy as previously reported by Cicman \textit{et al.} \cite{cicman1998dissociative}. 

The channel specific threshold energy value due to IPD process of CO$_2$ can also be determined with the help of known thermochemical parameters associated with the channel. Applying energy conservation principle all over the IPD process, we can write the following equation,

%Understanding the detailed dynamics and energetic responses occurring at the IPD region, i.e., beyond 15 eV incident electron energy is the centre of attention in this present work. In the following kinetic energy and angular distribution data taken out from velocity slice images (50 ns time sliced) of the anion formed due to the IPD process are analysed in depth. Fig.\ref{fig:ionyield} shows the ion yield curve around the threshold of the IPD region. Near the threshold energy value of ion yield, we do not get sharp-variation in anion formation due to poor energy resolution (0.6 - 0.8 eV) of our electron beam.

\begin{equation}
    \rm{V}_e = (E_i+D-A+IP)+E_1+E_2
\end{equation}
where V$_e$ is accounted for the energy transferred from the incident electron to the molecule, $E_i$'s are the energy associated with the possible excited states for the cation, D stands for the bond energy for the process; CO$_2 \longrightarrow $ CO + O, IP denotes the ionization potential energy for CO molecule and lastly A implies the electron affinity for oxygen atom. E$_1$ and E$_2$ denote the kinetic energies of the fragmented cation and anion, respectively. Again, E$_1$ and E$_2$ are restricted to zero at IPD threshold (V$_e=$ E$_{th}$). If we consider that both CO$^-$ and O$^-$ ions are produced on their corresponding accessible ground state i.e. E$_i=$0, then the threshold energy value of O$^-$ formation will obey the equation as below 
\begin{equation}
    \rm{(V_e)}_{threshold} = E_{th} = (D-A+IP)
\end{equation}
\begin{figure}[hbt]
    \centering
    \includegraphics[scale=0.34]{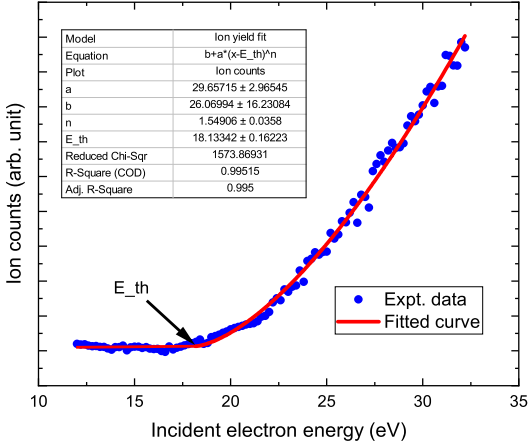}
    \caption{The data points in blue color represent the number of produced anion (O$^-$) as a function of electron beam energy, fitted using Fiegele \textit{et al.} \cite{fiegele2000threshold} model. The arrow in black color is used  to point out the threshold energy for anion formation.}
    \label{fig:threshold}
\end{figure} 
\begin{figure*}[hbt!]
    \centering
    \includegraphics[scale=0.33]{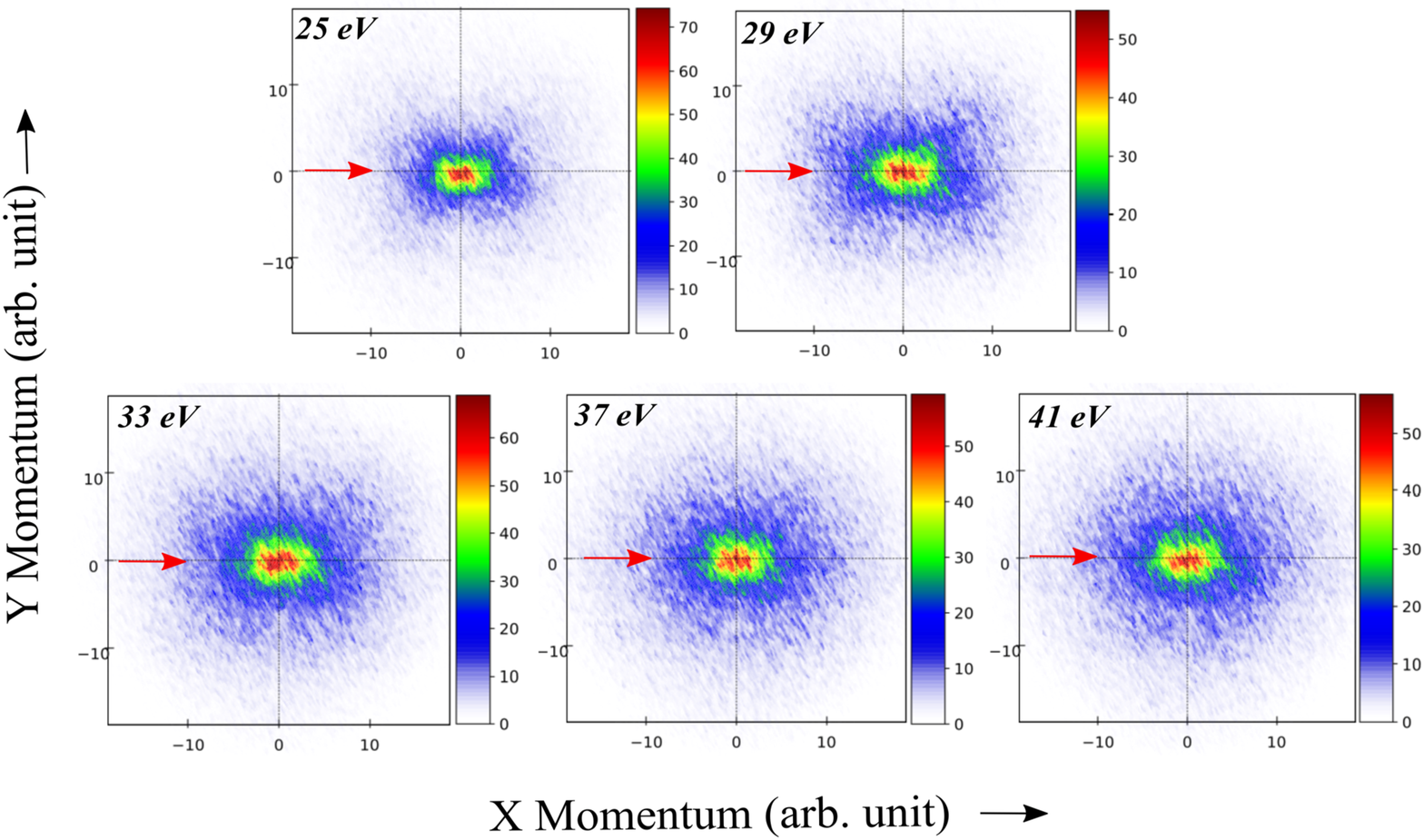}
    \caption{Time slices images taken with 50 ns time window of O$^-$ ions created due to the ion-pair formation process of CO$_2$ for five different incident electron energy. The arrow in red denotes the incident electron direction.}
    \label{fig:allvsi}
\end{figure*}

Now, if we use the known thermochemical values 5.51 eV (127 kcal/mol) \cite{darwent1970dissociation}, 1.46 eV \cite{blondel2001electron} and 14.014 eV \cite{erman1993direct} for the parameters D, A and IP, respectively; thermochemically obtained threshold energy value for the IPD process of CO$_2$ becomes 18.06 eV. Near threshold the experimental data points of the ion-yield curve are shaped using the model given by Fiegele \textit{et al.} \cite{fiegele2000threshold} and recently used by Nag \textit{et al.} \cite{nag2018study}.

\begin{align}
 \rm{f(E)} & = \rm{b \ \ \ \  for\  E<E_{th}}\\
      & = \rm{b\ + a(E-E_{th})^n \ \ \ for\ E>E_{th}}
\end{align}
Here, the parameter b is a constant appearing due to background correction. E$_{th}$ stands for the threshold energy value. The parameter a is the scaling factor that constraint to zero beneath the threshold. Lastly, n is the exponent which measures the slope of fitted curve after the threshold. Thus, n value is an indirect measurement of incident beam resolution. Due to finite energy resolution of incident electron beam the ion-yield curve increases slowly instead of sharp variation near the apparent threshold. To lessen this effect on threshold energy value estimation, we have fitted the data point 12.0 to 32.2 eV of incident electron energy using Eqn. 11 and the fit (best) gives it back more or less straight slope beyond the calculated threshold. The extrapolated straight line up-to threshold is nothing but the mean of background counts (parameter: b). The best fit shown by solid line in Fig. \ref{fig:threshold} is plotted using the values of different parameters as : a = 29.55715, b = 26.06994 and n = 1.549. Thus, we can say that our obtained threshold energy value (18.13 eV)  supports to a great extend in thermochemical view point.

During extraction of slice images, we have taken suitable ion-TOF (typically 50 nanoseconds) from the raw data in order to obtain central slice. Extracted time sliced velocity map images for fragmented anion (O$^-$) at 25 eV, 29 eV, 31 eV, 35 eV and 39 eV incident electron energies are shown in Fig. \ref {fig:allvsi}. The extracted slice-images are equivalent to that obtained by conventional methods using the inverse Abel transform, however, the artificial noise introduced by Abel transform can be avoided. The red arrow attached with each slice image indicates the direction of incident electron beam.

If we notice at the Fig. \ref{fig:allvsi} consisting with different slice images, it is clear to state that the maximum intensity of  O$^-$/CO$_2$ strikes near the centre part of sliced images and decreasing gradually with the increment the radius of 3D `Newton Sphere' i.e. most of the produced O$^-$ fragments are low energetic. Honestly, this type of two dimensional velocity slicing technique using time gate would take into account a larger fraction of the total solid angle for ions with lower momentum and a smaller fraction of solid angle for ions with larger momentum. In order to preserve the fraction of governed solid angle for any momentum to achieve better statistics, Moradmand \textit{et al.} \cite{moradmand2013momentum} defines a conical gate through wedge slicing technique instead of flat time gate; resulting the data restricts by angle in a spherical momentum-space, rather than in one Cartesian coordinate \cite{wu2012renner,slaughter2011dissociative}. As a counter response for this over-estimation in 2D time gated flat slicing technique, in a very recent article \cite{nag2019dissociative} Nag \textit{et al.} reported a new technique to study the characteristic of very low-energy ions. They have shown that an entire Newton sphere analysis brings the same results as canonical gated wedged slicing technique providing the similar nature of a single 0 eV peak in the kinetic energy distribution. However, during the study of sulfur dioxide using time gated velocity slice imaging technique Jana \textit{et al.} \cite{jana2019study} point out that although the 2D flat slicing method exaggerate the ions with lower momentum, the over-estimated part roughly consist of $~10^{-3}$ eV to the kinetic energy distribution. Thus, this does not change the underlying science qualitatively. In our case, we have taken the `Half Newton Sphere' to study the kinetic energy distribution instead of considering entire 'Newton Sphere' as the `Half Newton Sphere' also accounts equal fraction of solid angle for ions with all momenta. 
%Here, I have shown the pictorial representation for extracting the `Half Newton Sphere'.
%\begin{figure*}[hbt]
    %\centering
    %\includegraphics[scale=0.45]{slicingprocess.png}
    %\caption{pictorial representation for extracting the `Half Newton Sphere'}
    %\label{fig:my_label}
%\end{figure*}

\begin{figure}[hbt]
    \centering
    \includegraphics[scale=0.57]{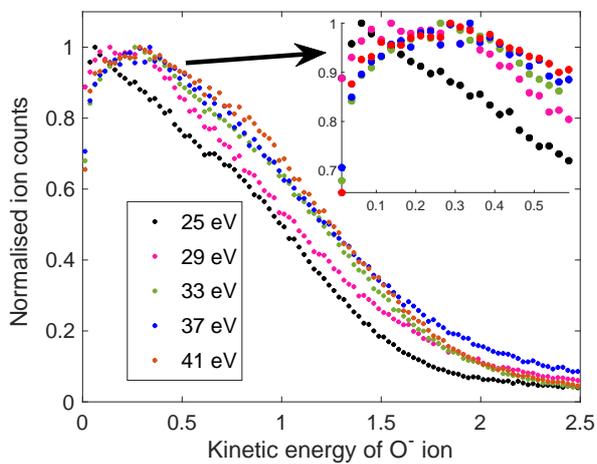}
    \caption{This plot shows the normalised kinetic energy distributions of the fragmented anion (O$^-$) for the five different beam energies. O$^-$ ions with lower momentum are shown in zoomed view at upper right corner of this plot. }
    \label{fig:kedist}
\end{figure}

We have aggregated all the time gated flat slices (each of 20 ns) extracted from the entire flight time for a particular ion where the average flight time of a slice act as bin-time of the corresponding slice. Then proper scaling of time-bin transfers the ellipsoidal shaped ion's packet into Newton sphere. Half Newton sphere is taken to consideration to obtain the kinetic energy distributions of the fragmented anion (O$^-$). The Fig. \ref{fig:kedist} shows the normalised kinetic energy distributions for the five different beam energies. The KE distribution shows maximum number of  O$^-$ ions are low energetic instead of zero in nature.
\begin{figure}
    \centering
    \includegraphics[scale=0.55]{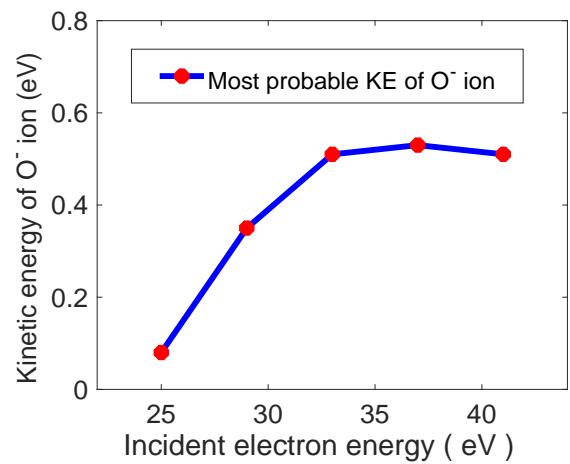}
    \caption{The most probable kinetic energy of the fragmented O$^-$ ions as a function of the incident beam energy. The most probable values are extracted using the Gaussian fit near the peak values.}
    \label{fig:variation}
\end{figure}

From the variation of most probable KE with incident electron energy as shown in Fig. \ref{fig:variation}, it is prominent that the ion kinetic energy first increases with the incident electron energy and then becomes locked. From previous reports \cite{chakraborty2016dipolar,nag2018study}, the formation of ion-pair dissociation can be explained due to presence of mainly two types of processes: direct and indirect process. In a direct process, the energy of the incident electron is transferred to the target molecule resulting in direct excitation of that neutral target to an ion-pair state. Such direct excitation dynamics are mainly governed by the Frank-Condon factor along with the valance-Rydberg mixing characteristics. Our obtained results in an increase in the anion kinetic energy (most probable value) with an increase incident electron energy; an irrefutable experimental indication of the direct transitions to the ion-pair state. However, In case of directly access ion-pair states, the variation of most probable kinetic energy of the fragmented ion-pair may governed by the partial energy transfer to the dissociation products.  During the study of electron impact dissociative ionization of ethylene Wang \textit{et al.} \cite{wang2015kinetic} obtains similar kind of mass depending average kinetic energy saturation phenomenon for different fragmented ions as a function of beam energies. on the other hand, in an indirect excitation the molecule may first get excited to a well defined Rydberg states (superexcited states) within the Frank-Condon region followed by transition to an ion-pair state through a predissociative cross-over; often occurs in case of excitation of molecules using photon. Such indirect excitation may result in a `locking' of the anion kinetic energy with increase in incident beam energy provided, the Rydberg and ion-pair states have same symmetry. Considering the kinetic energy distribution (Fig. \ref{fig:kedist}) of this present work, it can be seen that the ion KE first increases for incident electron energies 25, 29 and 33 eV; then becomes almost constant for 37 eV incident energy onward. The same can be noted from the most probable anion KE plot as shown in Fig. \ref{fig:variation}. It clearly noticeable that KE variations of the fragmented anion is small with respect to the variation of incident beam energies. Ion-pair state achieved directly via. electron collision may transfer its energy partially to dissociate the molecule; could be a rationale for such KE variation of fragmented anion. Beyond 33 eV incident electron energy, the most probable anion KE becomes saturated due to unavailability for accessing the ion-pair state directly. However, ion-pair generation via. direct access to the ion-pair state is strongly depends on molecular geometry; where as indirectly produced ion-pair is a generalization for all target molecules.

\begin{figure}[hbt]
    \centering
    \includegraphics[scale=0.35]{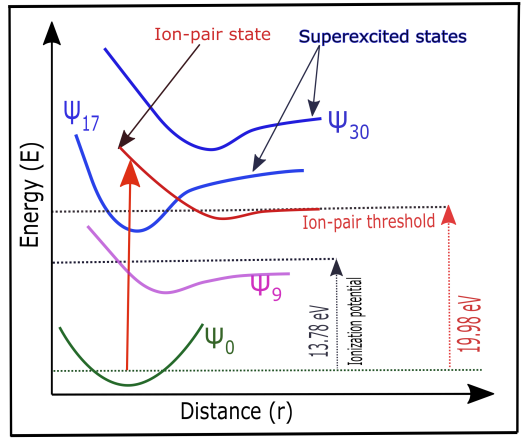}
    \caption{Potential energy curves of CO$_2$ along with ion-pair state in red line are shown schematically. Among the calculated nonzero oscillatory strength, $\Psi_{17}$ and $\Psi_{30}$ states positioned above the ionization potential energy of CO$_2$ i.e. superexcited states shown in blue line. The ordinary Rydberg state ($\Psi_{9}$) in pink line lies well below the ionization potential \cite{tanaka1960higher}. Red vertical arrow using solid line indicates the direct transition to the ion-pair state.}
    \label{fig:schematic}
\end{figure}

\begin{table}[hbt]
 \centering
\caption{state specific characteristics of CO$_2$ molecule obtained from DFT and TDDFT calucations.}  
\begin{tabular}{cccc}

\hline
\hline
%\multicolumn{2}{c}{Item} \\
%\cline{1-2}
    \textbf{\vtop{\hbox{\strut \ \ \ Energy }\hbox{\strut eigenvalue }}\hspace{2mm}}    & \textbf{\vtop{\hbox{\strut Total energy}\hbox{\strut \ \ \ \ \ in eV}\hbox{\strut }}\hspace{2mm}}    & \textbf{\vtop{\hbox{\strut Symmetry}\hbox{\strut attached}}\hspace{2mm}}  & \textbf{\vtop{\hbox{\strut Oscillatory}\hbox{\strut \ strength}}\hspace{2mm}}\\
\hline

Ground & 00.00 & $\Sigma_g$ & NA \\
%\hline

9$^{th}$ & 11.31 & $\Sigma_u$ & 0.343 \\ 
%\hline

17$^{th}$  & 16.33 & $\Sigma_u$ & 1.498 \\ 
%\hline

30$^{th}$  & 24.19 & $\Sigma_u$ & 0.207 \\ 
%\hline
\hline
\hline
\end{tabular}
\end{table}
However, considering the quantum chemistry computation calculated using the above mention framework, the 17$^{th}$ excited state seem to have the maximum contribution followed by the 30$^{th}$ excited state (as guided by the decreasing non-zero oscillatory strength values). Again, we may conclude that significant amount of the ion-pair dissociation results from a direct transition to the ion-pair excited state. To give a clear idea of the above mentioned processes, we have given a schematic of the potential energy curves for CO$_2$ as shown in Fig. \ref{fig:schematic}. A firm conclusion delineating the direct and indirect excitations would require a nuclear dynamics calculation which is beyond the scope of this work. Also, we have calculated total ground states energies of CO$^+$ and O$^-$ ions using same level of theory. The minimum energy required for transition directly from CO$_2$'s ground to these ionic ground states is about 19.98 eV, not distinctly far away from our experimental observation in appearance of O$^-$ ion. 

%In Fig.\ref{fig:schematic}, we have schematically positioned  ion-pair states beyond the Frank-Condon transition factor to obliterate the direct one. In a similar fashion, the 17$^{th}$ energy eigenvalue has greater Frank-Condon factor with ground state of CO$_2$ as comparison to 30$^{th}$ energy eigenvalue. 
\begin{figure}[hbt!]
    \centering
    \includegraphics[scale=0.25]{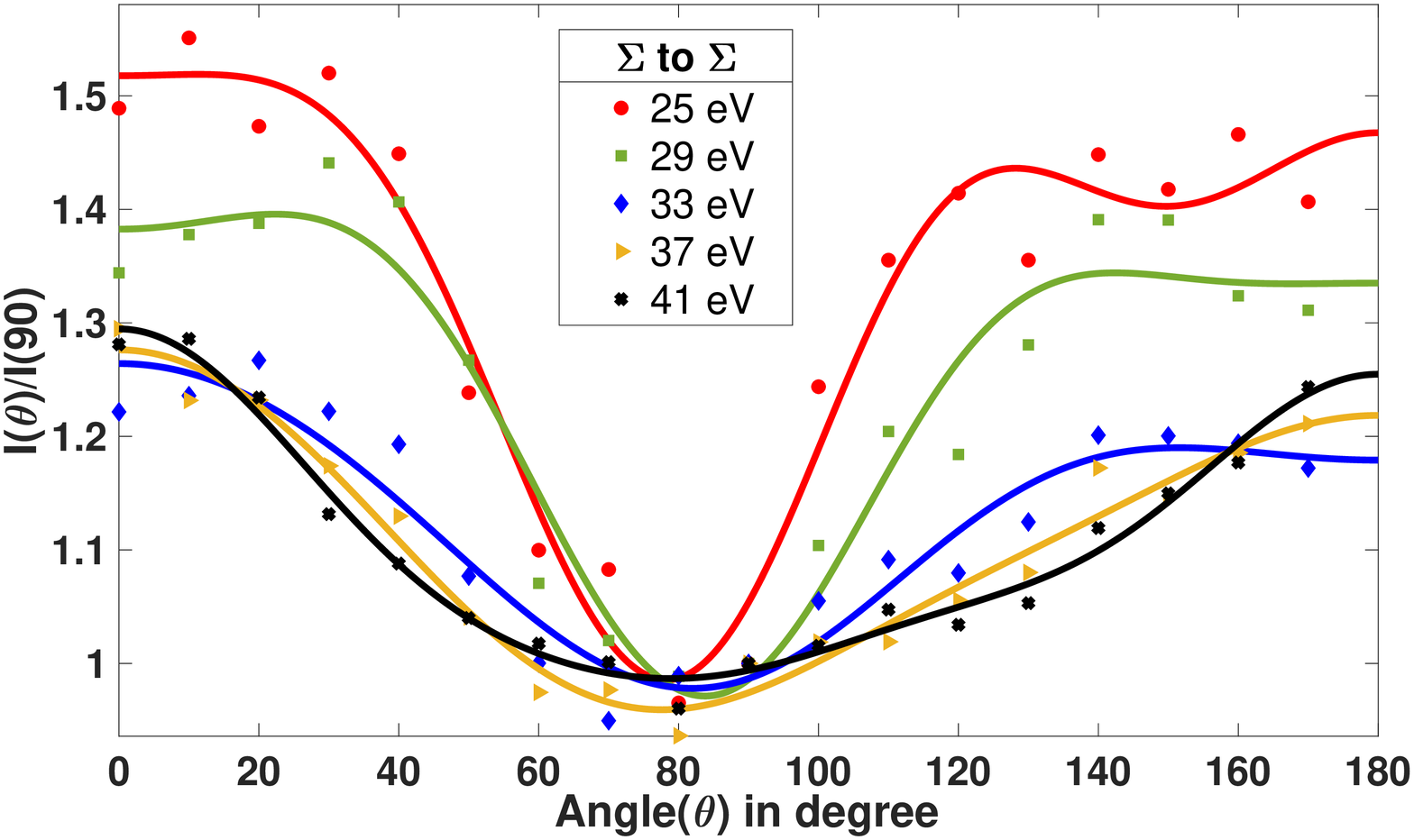}
    \caption{Figure shows the angular distribution of the fragmented anions (O$^-$) due to the IPD process for CO$_2$. The solid lines in different color are fit to the data for $\Sigma$ final state only.}
    \label{fig:ss}
\end{figure}
\begin{figure}[bht!]
    \centering
    \includegraphics[scale=0.25]{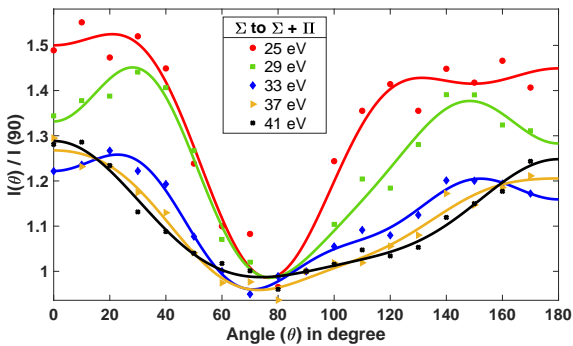}
    \caption{Figure shows the angular distribution of the fragmented anions (O$^-$) due to the IPD process for CO$_2$. The solid lines in different color are fit to the data for $\Sigma + \Pi$ final states.}
    \label{fig:ssp}
\end{figure}

Angular distributions study of the fragmented ion-pairs contributes a significant determinant to gauge the molecular symmetry associated with the ion pair state. The angular distributions of produced O$^-$ ions as shown in Fig. \ref{fig:ss} and  Fig. \ref{fig:ssp}, have been analysed for anion kinetic energy from 0 to 2.5 eV. During the AD study, 50 nanoseconds slices of O$^-$/CO$_2$ have been taken into account. Due to experimental difficulties of determine simultaneously both the distribution of the fragments and the scattering angles of the electrons, we have set the direction of electron beam as a convenient reference system for reporting the form of angular distribution. All the data points have been normalised at 90$^{\circ}$. In Fig. \ref{fig:ss} and Fig. \ref{fig:ssp} symbols represents the experimentally obtained data points and the solid curves are the fit-to data using the expression given by Van Brunt \cite{van1974breakdown}.

\begin{table*}[bht!]
 \centering
\caption{The model parameters reported here for the O$^-$/CO$_2$ angular distribution as a function of electron energies with contributions from $\Sigma$ final state only. The a$_m$’s are the relative strengths of various partial waves for contributing to the $\Sigma$ final states. The $\delta_l$’s expressed in radian signify the phase differences with respect to the lowest order partial wave responsible for the transition. $\beta_1$’s are the product of the momentum transfer vector and the distance of closest approach between the incident electron and the centre of mass of the molecule for $\Sigma$ final state only. Lastly, each m used in this caption is nothing but the whole numbers.}  
\begin{tabular}{cccccc}
\hline
\hline
%\multicolumn{2}{c}{Item} \\
%\cline{1-2}
     \textbf{\vtop{\hbox{\strut Incident electron }\hbox{\strut \ \ \   energy (eV) }}\hspace{5mm}}    & \textbf{\vtop{\hbox{\strut Weighting ratio of the}\hbox{\strut different partial waves}\hbox{\strut $\hspace{8mm}a_0:a_1:a_2:a_3$}}\hspace{5mm}}    & \textbf{\vtop{\hbox{\strut \ \ \ \ \  Phase differences in radian }\hbox{\strut \ \ }\hbox{\strut \ \ \ \ \ \ \ \ \ \ \ \ \ \  $\delta_{s-p}$, $\delta_{s-d}$, $\delta_{s-f}$}}\hspace{5mm}}  & \textbf{\vtop{\hbox{\strut$\beta$ parameter}\hbox{\strut }\hbox{\strut \ \ \ \ \ \ \ \ \ $\beta_1$}}\hspace{5mm}}   & \textbf{\vtop{\hbox{\strut$R^{2}$ value}\hbox{\strut }}\hspace{5mm}}\\
\hline

25     &  1.94 : 1.16 : 2.17 : 8.65     & 2.99, 1.75, 3.14   & 1.00 & 0.940 \\
%\hline

29     &  1.65 : 1.74 : 3$\times 10^{-9}$ : 9.71     & 0.16, 3.14, 0.21   & 0.79 & 0.897\\
%\hline

33     & 1.54 : 1.59 : 2$\times 10^{-5}$ : 9.08   & 0.02, 0.71, 0.50   & 0.58 & 0.904 \\
%\hline

37     &  1.59 : 0.02 : 2.44 : 7.83     &1.66, 2.07, 2.67   & 0.60 & 0.964 \\
%\hline

41    & 1.52 : 0.26 : 2.29 : 4.88    & 0.02, 3.14, 1.93     & 0.44 & 0.938\\
\hline
\hline
\end{tabular}
\end{table*}

\begin{table*}[hbt]
 \centering
\caption{The model parameters reported here for the O$^-$/CO$_2$ angular distribution as a function of electron energies with contributions from $\Sigma + \Pi$ final states. The a$_m$’s and b$_m$’s are the relative strengths of various partial waves for contributing to the $\Sigma$ and $\Pi$ final states, respectively. The $\delta_l$’s expressed in radian signify the phase differences with respect to the lowest order partial wave responsible for the transition. $\beta_1$’s and $\beta_2$’sare the product of the momentum transfer vector and the distance of closest approach between the incident electron and the centre of mass of the molecule for $\Sigma$ and $\Pi$ final states, respectively. Lastly, each m used in this caption is nothing but the whole numbers.}
\vspace{2mm}
\begin{tabular}{cccccc}

\hline
\hline
%\multicolumn{2}{c}{Item} \\
%\cline{1-2}
   \textbf{\vtop{\hbox{\strut Incident electron }\hbox{\strut \ \ \ \ energy (eV) }}\hspace{5mm}}    & \textbf{\vtop{\hbox{\strut Weighting ratio of the}\hbox{\strut different partial waves}\hbox{\strut $\hspace{10mm}a_0 : a_1 : a_2 : a_3$}\hbox{\strut $\hspace{10mm}b_1 : b_2  : b_3 : b_4$}}\hspace{5mm}}    & \textbf{\vtop{\hbox{\strut \ \ \ \ \  Phase differences in radian}\hbox{\strut \ }\hbox{\strut \ \ \ \ \ \ \ \ \ \ \ \ \  $\delta_{s-p}$, $\delta_{s-d}$, $\delta_{s-f}$}\hbox{\strut \ \ \ \ \ \ \ \ \ \ \ \ \  $\delta_{p-d}$, $\delta_{p-f}$, $\delta_{p-g}$}}\hspace{5mm}}  & \textbf{\vtop{\hbox{\strut$\beta$ parameter}\hbox{\strut }\hbox{\strut \ \ \ \ \ \ \ \ \ $\beta_1$}\hbox{\strut \ \ \ \ \ \ \ \ \ $\beta_2$}}\hspace{5mm}}   & \textbf{\vtop{\hbox{\strut$R^{2}$ value}\hbox{\strut }}\hspace{5mm}}\\
\hline
%\multirow{1}{*}{Season} & \multicolumn{1}{c}{Lower Limit} & \multicolumn{1}{c}{Upper Limit} & \multicolumn{1}{c}{Upper Limit} & \multicolumn{1}{c}{Upper Limit} \\

\multirow{2}{*}{25 }   &1.37 : 2.50 : 3.08 : 6.54    & 4.8$\times 10^{-4}$, 4.3$\times 10^{-14}$, 0.75  & 1.081 & \multirow{2}{*}{0.943}\\
%\cmidrule
                       & 0.03 : 0.62 : 0.72 : 1.25   & 0.20, 1.25, 1.24       & 0.058 & \\
\hline

\multirow{2}{*}{29 }   &1.15 : 0.10 : 10.88 :7.48   & 0.06, 3.14, 1.38       & 0.409 & \multirow{2}{*}{0.975}\\
                       &1.87 : 0.01 : 5.13 : 8.73    & 0.20, 2.01, 2.55       & 1.418 & \\
\hline

\multirow{2}{*}{33 }   &1.51 : 1.48 : 0.18 : 3.04   & 0.01, 0.20, 1.00    & 0.685 & \multirow{2}{*}{0.985}\\
                       &0.27 : 1.21 : 4.21 : 2.73        & 0.38, 0.08, 2.40       & 1.450 & \\
\hline

\multirow{2}{*}{37 }   &1.64 : 0.03 : 1.75 : 4.95   & 0.50, 2.08, 2.66  & 0.702 & \multirow{2}{*}{0.964}\\
                       & 0.006 : 0.047 : 0.15 : 4.839       & 0.035, 2.365, 1.98       & 0.008 & \\
\hline

\multirow{2}{*}{41 }   &1.56 : 0.075 : 5.42 : 4.70    & 0.005, 1.75, 2.28  & 0.559 & \multirow{2}{*}{0.978}\\
                         & 0.126 : 5.84 : 0.27 : 0.274      & 1.89, 2.31, 0.236       & 0.059 & \\
\hline
\hline
\end{tabular}
\end{table*}

\begin{equation}
    \rm{I}(\theta)=K^{-n} \left\lvert \sum_{l = \mid \mu \mid}^{\infty} i^{l} \sqrt{\frac{(2l+1)(l-\mu)!}{(l+\mu)!}} j_{l}(\beta) Y_{l, \mu}(\theta, \phi)\right\rvert^{2}
\end{equation}

Here, K is the momentum transfer vector between the incident and scattered electron, j$_{l}$'s are the spherical Bessel function, $\beta$ denotes the product of the momentum transfer vector and the distance of closest approach between the incident electron and the centre of mass of the molecule, Y$_{l, \mu}$'s are the spherical harmonics, \textit{l} is the angular momentum of the electron and $\mu=\mid \Lambda_f - \Lambda_{i}\mid$, where $\Lambda_{i}$ and $\Lambda_{f}$ are the projection of angular momentum along the molecular axis for the initial and final states, respectively. Here, the summation over \textit{l} is responsible for the attachment of different partial waves. However, the transition occurs between two specified states for a particular incident electron's energy then values K and n would be fixed and  can be treated as parameters. So, the AD data due to the inclusion of one or more than one final states must be fitted using the expression  
\begin{equation}
    \rm{I}(\theta) = \sum_{\mid \mu \mid }\left\lvert \sum_{l = \mid \mu \mid}^{\infty} a_{l}i^{l} \sqrt{\frac{(2l+1)(l-\mu)!}{(l+\mu)!}}
    j_{l}(\beta) Y_{l, \mu}(\theta, \phi)e^{i\delta_l}\right\lvert^{2}
\end{equation}

Where a$_l$'s are the energy dependent weight factors for different partial waves and $\delta_l$'s are considered to calculate the phase difference of that particular transitional partial wave with respect to the lowest order possible partial wave. The summation over $\mu$ signifies the contribution of more than one ion-pair states to this process. During the AD fitting, we have considered possible lowest four partial waves for both $\Sigma$ to  $\Sigma$  and $\Sigma$ to  $\Sigma + \Pi$ transitions.

Being attached with the electron, temporary negative ion (TNI) of CO$_2$ departs from linear geometry before dissociation \cite{moradmand2013dissociative,jana2019probing} and declines to obey axial recoil approximation. In case of ion-pair dissociation, electron attachment to the molecule is no longer a valid statement. Thus, considering axial recoil as a valid statement we can demand the symmetry attached to the super-excited states would preserve in ion-pair states also. We have already tabulated the symmetry and transition probability of different states of CO$_2$ while demonstrating the KE distribution of the fragmented anions. Here the ground state symmetry of CO$_2$ i.e. $\Sigma$ is considered as initial one for all of the involved transitions. In Fig. \ref{fig:ss} and Fig. \ref{fig:ssp} the solid lines in the AD distribution represent the IPD dynamics with symmetry $\Sigma$ and $\Sigma + \Pi$, respectively, for all of the final states. However, the goodness (R$^2$ value) for curve fitting to $\Sigma + \Pi$ final states is slightly better than $\Sigma$ final states. Theoretically, we have not obtained any $\Pi$ symmetry involvement in the super-excited states. \\

In Fig. \ref{fig:ss} and Fig. \ref{fig:ssp} it is clearly noticeable that number of scattered O$^{-}$ ions  in forward direction is slightly greater than scattered in  backward direction. Misakian \textit{et al.} \cite{misakian1972linear} reported that this lack of symmetry about $\theta=$90$^{\circ}$ aries due to the linear momentum transfer effect consideration during the laboratory frame angular distribution measurement; whereas thermal molecular motions alone are found to have no effect on the center of mass angular distribution. Though, this asymmetry in lab frame angular distribution is prominent near threshold for lighter molecules. During the study of H$_2$O through DEA resonance Haxton \textit{et al.} \cite{haxton2006angular}  and Ram \textit{et al.} \cite{ram2009resonances} point out that forward backward asymmetry in AD of dissociative products aries due to the contribution of partial waves in different weight with different phase. Moreover, the momentum transfer (either linear or angular) plays a crucial role to design such maintained asymmetry.
\begin{figure}[htb!]
    \centering
    \includegraphics[scale=2.2]{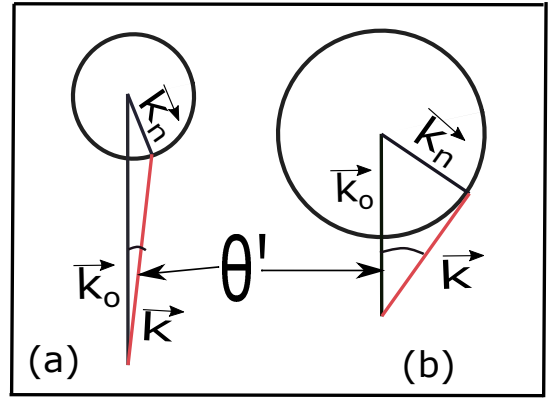}
    \caption{This schematic represents the possible orientation of ${(\vec{\rm{k_o}})}$, $(\vec{\rm{k_n}})$ and $(\vec{\rm{k}})$ as previously discussed by Zare \cite{zare1967dissociation}, (a) although $\rm{(\theta^{'})}$ and $(\vec{\rm{k}})$ are small near IPD threshold, contribute heavily to the form of $\rm{I}(\theta)$, (b) bigger angle of $\rm{(\theta^{'})}$ at far from IPD threshold, makes $(\vec{\rm{k}})$ to weight loosely in $\rm{I}(\theta)$.}
    \label{fig:zare}
\end{figure}

Interestingly, at higher incident electron energies the angular distribution gradually losses it's anisotropic nature. This reduction of anisotropy in IPD dynamics could be explained by much the same argument given by Zare \cite{zare1967dissociation}. During the theoretical study of angular distribution of the  H$^{+}_2$ ions from electron impact dissociation Zare reported that the nature of AD of dissociation product, calculated with respect to the electron beam direction, depends on the angle $\rm{(\theta^{'})}$ generated by wave vector ${(\vec{\rm{k_o}})}$ along the incident electron beam direction and the direction of wave vector $(\vec{\rm{k}})$ responsible for the momentum transfer to the molecule. Near threshold this angle $(\theta^{'})$ is restricted to small and increases gradually with the increment of  incident beam energies as shown in Fig. \ref{fig:zare}; resulting the lab frame distribution disfigures from a cosine-squared dependence for $(\vec{\rm{k}})$ parallel to $(\vec{\rm{k_o}})$  to a sine squared dependence for $(\vec{\rm{k}})$ perpendicular to $(\vec{\rm{k_o}})$. Here, $(\vec{\rm{k_n}})$ denotes the wave vector corresponding to the n$^{th}$ transitional state of that molecule. Thus, it is not like that the absolute value of momentum transfer vector is decreasing with the increment of incident beam energies, it is due to the decreasing nature of the momentum transfer weight factor to the intensity distribution. If we focus on the $\beta$  parametric values for $\Sigma$ final states, it is clear that cosine-squared dependency to AD is diminishing gradually as a function incident beam energies. Thus, in IPD dynamics of CO$_2$, the momentum transfer vector weights strongly to the differential cross section for lower beam energies. Under the restriction imposed for contributing from the higher order partial waves $(l>1)$ to the angular distribution, Zare also predicts that if the measurements of Dunn and Kieffer \cite{dunn1963dissociative} were extended towards threshold, the fragment distribution would not approach a "pure" cosine-squared anisotropy, but would resemble like our obtained angular distribution shown in Fig. \ref{fig:ss} and Fig. \ref{fig:ssp}.

\section{Conclusions}
Here, a quite similar process of ion-pair imaging spectroscopy (IPIS) has been sucessfully applied in case of low energy electron-molecule interaction. Determination of the threshold energy value to initiate the ion-pair formation matches well with the thermochemically obtained value. Here, the dissociation dynamics as well as the product of dissociation reported through out this analysis is based on the fragmented anion characteristic only; coincidentally measurement of both the anion and cation products can overstoke our findings. Detail calculations for the ion-pair generation may provide an explanation for the surprisingly strong direct process; a difficult job to the scientific community. However, Krauss \textit{ et al.} \cite{krauss1975ion} have reported this valance-Rydberg mixing is relatively small in case of oxygen. We have assigned the symmetries to the ion-pair states based on the angular distribution fit using equation given by Van Brunt \cite{van1974breakdown}. Ion-pair states associated with $\Sigma$ and $\Pi$ symmetry have been identified throughout the entire kinetic energy range (0 - 2.5 eV) of produced anion; where as the contribution from $\Pi$ symmetry is notably low in comparison with $\Sigma$ symmetry. Too much cosine-squared dependency in AD could be a rationale for the dominance of such $\Sigma$ symmetric ion-pair states. Interestingly, we have still hearken with two dimensional time gated flat slicing technique to analyse the AD instead of three dimensional method; as slices out of the incident electron beam plane territory will make an angle with the incident beam, resulting diversion from the in general anisotropic nature of dissociation products. Thus, our referencing direction (electron beam direction) for obtaining AD should be redefined for considering three dimensional Newton sphere (either half or full) analysis. Due to unavailability of complete theoretical demonstration of IPD dynamics, state forward determination of the symmetry associated with a particular ion-pair state is a challenging job to the theoretical science community. However, our density functional theory (DFT) calculations for CO$_2$ Rydberg states enable us to define its super excited states and also provide the symmetries associated with the ion-pair states. Nevertheless, in case of IPD dynamics induced by electron impact on carbon dioxide, straightforward dipole interpretation is no longer carried out during angular distribution measurement of produced anions.

\section{Acknowledgements}

N. K. gratefully acknowledges the financial support from `DST of India' for "INSPIRE Fellowship" program. A. P. express deep appreciation to "Council of Scientific and Industrial Research (CSIR)" for the financial assistance. We gratefully acknowledge financial support from the Science and Engineering Research Board (SERB) for supporting this research under  Project No. “ EMR/2014/000457 ”.

\bibliography{mybibfile}

\end{document}